# SUPERHUMPS IN CATACLYSMIC BINARIES.

# XXII. 1RXS J232953.9+062814

DAVID R. SKILLMAN[1], THOMAS KRAJCI[2], EDWARD BESHORE[3],

JOSEPH PATTERSON[4], JONATHAN KEMP[5,4], DONN STARKEY[6], ARTO OKSANEN[7],

TONNY VANMUNSTER[8], BRIAN MARTIN[9], AND ROBERT REA[10]

[1] Center for Backyard Astrophysics (East), 9517 Washington Avenue, Laurel, MD 20723; dskillman@home.com

[2] Center for Backyard Astrophysics (New Mexico), 1688 Cross Bow Circle, Clovis, NM 88101; krajcit@3lefties.com

[3] Center for Backyard Astrophysics (Colorado), 14795 East Coachman Drive, Colorado Springs, CO 80908; ebeshore@pointsource.com

[4] Department of Astronomy, Columbia University, 550 West 120th Street, New York, NY 10027; jop@astro.columbia.edu

[5] Joint Astronomy Centre, University Park, 660 North A`ohōkū Place, Hilo, HI 96720; j.kemp@jach.hawaii.edu

[6] Center for Backyard Astrophysics (Indiana), 2507 County Road 60, Auburn, IN 46706; starkey@fwi.com

[7] Center for Backyard Astrophysics (Finland), Vertaalantie 449, Nyrölä, Finland; arto.oksanen@jklsirius.fi

[8] Center for Backyard Astrophysics (Belgium), Walhostraat 1A, B–3401 Landen, Belgium; tonny.vanmunster@advalvas.be

[9] King's University College, Department of Physics, 9125 50th Street, Edmonton, AB T5H 2M1, Canada; bmartin@kingsu.ab.ca

[10] Center for Backyard Astrophysics (Nelson), 8 Regent Lane, Richmond, Nelson, New Zealand; reamarsh@ihug.co.nz

## ABSTRACT

We report photometry of a recently discovered dwarf nova with a remarkably short 64.2-minute orbital period. In quiescence, the star's light curve is that of a double sinusoid, arising from the "ellipsoidal" distortion of the Roche-lobe-filling secondary. During superoutburst, common superhumps develop with a period 3–4% longer than $P_{orb}$. This indicates a mass ratio $M_2/M_1$=0.19±0.02, a surprisingly large value in so compact a binary. This implies that the secondary star has a density 2–3 times higher than that of other short-period dwarf novae, suggesting a secondary enriched by H-burning prior to the common-envelope phase of evolution. We estimate $i$=50±5°, $M_1=0.63^{+0.12}_{-0.09}$ $M_{\odot}$, $M_2=0.12^{+0.03}_{-0.02}$ $M_{\odot}$, $R_2=0.121^{+0.010}_{-0.007}$ $R_{\odot}$, and a distance to the binary of 180±40 pc.

## 1. INTRODUCTION

The star 1RXS J232953.9+062814 (hereafter RX 2329+06) was discovered as an X-ray source by *ROSAT*, and identified as a cataclysmic variable by Jingyao et al. (1998). In November 2001 the star jumped into outburst (Schmeer 2001), and the first night of time-series photometry showed periodic waves with a period near 66 minutes (Uemura et al. 2001). We started a program of time-series photometry immediately, and steadily tracked the star over the next 53 days.

The binary is remarkable for its very short orbital period of 64.1766(3) minutes (Thorstensen et al. 2002, hereafter T02), well below the famous "period minimum" for hydrogen-rich secondaries. This implies that the secondary is smaller than a hydrogen-rich secondary, a conclusion strongly supported by analysis of the periodic waves in the light curve. A preliminary account of our photometric campaign has already appeared in T02; here we give full details.

## 2. OBSERVATIONS

We began photometric observations on JD 2452219, two nights after Schmeer's outburst discovery. We used our network of backyard telescopes (the Center for Backyard Astrophysics, or CBA) to obtain differential photometry, and spliced the nightly light curves together to obtain long time series and hence accurate periods. Most of the photometry was unfiltered, but some was obtained through standard *UBVRI* filters. Skillman & Patterson (1993) describe the observation and analysis techniques. The geographical distribution of telescopes enabled us to measure periods accurately without aliasing. Coverage was obtained on 42 of 53 nights during 2001–2, totaling 285 hours. The participating observers and telescopes are listed in Table 1.

### *2.1 LIGHT CURVES AND PERIODS*

We formed an eruption light curve by measuring the average magnitude from each night's observation (or occasionally two, when severe trends were present). This is shown in Figure 1. The eruption could have started anywhere in the window JD 2452210–6; so our observations started relatively late in the outburst. In this paper we shall use truncated dates: JD = true JD – 2452200. On JD 21.6, the star dipped briefly to $V$=15.5, recovered to a second brief outburst at JD 23.8 (sometimes called an "echo" outburst), and then faded monotonically to quiescence at $V$=16.5. A very short eruption, lasting <2 days, occurred on JD 70. Time-series photometry showed prominent "superhumps" in the first, much longer outburst — and no significant modulations in the short outburst near the end of the season's coverage. In accordance with common practice, we therefore identify the first as a *superoutburst*, and the last as a *normal* outburst.

Several nightly light curves during superoutburst are shown in the upper frames of Figure 2, illustrating the prominent 66-minute waves. The lower frame shows a power spectrum over the first 6 nights of coverage, with a mean light curve at $P$=0.0463 d inset. The association with superoutburst (and the discrepancy with $P_{orb}$, discussed by T02) establishes that these are superhumps.

We divided the data into several-day intervals in order to follow the evolution of the periodic signals. The frequencies detected in the power spectra are given in Table 2. The stable signal at 44.877 c/d is merely the orbital signature at $2\omega_{orb}$, discussed by T02. The waves prominent in superoutburst clearly change in frequency, with the fundamental ranging from 21.57 to 21.80 c/d.

### *2.2 SUPERHUMP EVOLUTION*

This matter of a changing superhump frequency merits closer study. A sensitive measure of period change is provided by the O–C diagram, which tracks phase changes from a constant-period ephemeris. We timed all the moments of superhump maxima, and present them in Table 3. (Timings after JD 28 were calculated after subtracting the orbital waveform, and averaging over the night's data.) Then we compared to the test ephemeris HJD 19.361+0.046*E*. The result, seen in the lower frame of Figure 3, shows a decreasing period, consistent with the frequencies in Table 2. The early phase of the superhump, which we have labelled the *common* superhump, is described by

$$\text{Maximum light} = \text{HJD } 19.361 + 0.04654\, E - 0.0000016\, E^2.$$

This corresponds to a period change given by $\dot{P}/P = 5\times10^{-5}$.

On JD 29, or cycle 200, the superhump showed a rapid phase change. Figure 3 shows a jump of –0.38±0.07 (or +0.62±0.07) cycles. This is probably an example of a "late superhump", a fairly standard phenomenon in well-observed dwarf novae (van der Woerd et al. 1988, Hessman et al. 1992). The superhump became undetectable (falling below 0.04 mag full amplitude) after JD 42.

The upper frame of Figure 3 shows the light curve during the superhump evolution. It is notable that the echo outburst has no effect on superhump phase (and the trend in amplitudes, not shown here, also shows a superhump constant in flux units throughout the echo). Also notable is the long duration of the superhump, at least 400 cycles after the main outburst ended on JD 22.

## 3. THE ORBITAL SIGNAL

As the outburst light subsided on JD 25, a powerful signal at 44.88 c/day started to appear in the light curve. The period during JD 28–68 is 0.022283(2) d, exactly half the orbital period revealed by the radial-velocity variations (T02). The power spectrum of JD 29–68 is shown in Figure 4, with the inset figure showing the mean waveform at $P_{orb}$. The latter is a "double-sinusoid" with a measured full amplitude (the average difference between maxima and minima) of 0.14±0.02 mag, and minima differing by 0.030±0.012 mag. This signal was apparently also present just before superoutburst, in the photometry reported by Zharikov & Tovmassian (2001).

The signal reaches primary minimum at orbital phase 0.51±0.01, just as the secondary star reaches superior conjunction in the T02 ephemeris. This establishes an origin in the "ellipsoidal" variations of the Roche-lobe-filling secondary (T02). Correcting for the presence of unmodulated disk light, we estimate that the secondary's intrinsic amplitude is $A$=0.20±0.03 mag, with minima asymmetric by $\Delta A$=0.036±0.014 mag.

Unfortunately, we cannot use the measured value of $\Delta A$ for further analysis, because accretion disks typically produce photometric signals at $P_{\rm orb}$, from causes unrelated to the distortion of the secondary. So we shall have to settle for just the constraint on $A$.

The theory for the interpretation of such light curves has been treated by Bochkarev, Karitskaya, & Shakura (1979, hereafter BKS). BKS discuss the dependence of $A$ on $q$, the binary inclination $i$, the limb-darkening coefficient $u$, and the gravity-darkening coefficient $\beta$. For the K-star secondary at ~6000 Å, appropriate choices are $u$=0.6 and $\beta$=0.5 (BKS, Binnendijk 1974). Interpolation in Table 1 of BKS then yields the observed range of $A$ for the relevant $q$ (see below) when $i$=50±5°.

## 4. PROPERTIES OF THE BINARY

### *4.1 MASS RATIO*

Since RX 2329+06 shows two sets of lines, the spectroscopy of T02 furnishes a direct formal solution for $q=K_1/K_2$=0.21±0.02. However, T02 reported a $K_1$ phase shift of 0.05 cycles relative to the fiducial phase of the white dwarf (exactly opposite the secondary), which makes it not quite reliable in a dynamical argument. So we seek another constraint. The superhump period gives an independent measure of $q$, since the observed fractional period excess ε is approximately proportional to $q$ (in theory, and evidently also in practice; Patterson 2001). For the main part of the superoutburst, RX 2329+06 showed ε=0.037±0.003, which implies $q$=0.17±0.03 according to Patterson's Eq. (5). Combining the constraints, we estimate $q$=0.19±0.02.

### *4.2 MASSES*

RX 2329+06 is essentially a non-eclipsing single-lined binary, and hence fails by *two* constraints to yield masses directly. The curves in Figure 5 show the solutions for the spectroscopic mass function of 0.194±0.006 $M_\odot$ (T02). We can add the $q$ constraint cited above, and the $i$ constraint provided by the ellipsoidal variation. The result is the black region in Figure 5, with masses

$$M_2 = 0.12^{+0.03}_{-0.02}\ M_\odot \qquad (2)$$
$$M_1 = 0.63^{+0.12}_{-0.09}\ M_\odot.$$

### *4.3 NATURE OF THE SECONDARY*

Secondary stars in CVs fill their Roche lobes, which constrains the secondary's mass $M_2$ and radius $R_2$:

$$P_{\rm orb}\,[\mathrm{hr}] = 8.75\,(M_2/R_2^{\,3})^{-1/2}, \quad (3)$$

with $M_2$ and $R_2$ in solar units (Faulkner, Flannery, & Warner 1972). For this binary we deduce

$$M_2/R_2^{\,3} = 66.9. \quad (4)$$

Thus for our secondary of $0.12^{+0.03}_{-0.02}\ M_\odot$, we deduce $R_2$=$0.121^{+0.010}_{-0.007}\ R_\odot$.

For the observed K5 spectral type (T02), stars in the field have $M$~0.7 $M_\odot$, $R$~0.7 $R_\odot$. So of course the secondary is *vastly* smaller than expected for its spectral type. Even ignoring the spectral type, it is still too small — at 0.12 $M_\odot$, a zero-age main sequence star of cosmic composition has $R$=0.15 $R_\odot$ (Baraffe et al. 1998), and the lobe-filling secondaries of CVs appear to have $R$≈0.17 $R_\odot$ (see Figure 2 of Patterson 2001). The secondary in this binary is too hot and too small to be consistent with an unevolved star [T02; see also Augusteijn et al. (1996) for a discussion of V485 Centauri, likely a close relative].

Since we know the radius and $T_{\rm eff}$ of the secondary, we can estimate its absolute magnitude $M_V$=10.6±0.2 (scaling from K dwarfs in the Hyades). Comparison with the observed $V$=16.9±0.2 yields a distance estimate of 180±40 pc.

## 5. SUMMARY

1. Photometry of RX 2329+06 in superoutburst reveals powerful waves with a period averaging 66.4 minutes. These appear to be garden-variety common superhumps, decreasing slightly in period with $\dot{P}/P$=5×10$^{-5}$. There was probably a transition to "late" superhumps, and the entire superhump episode ended ~400 cycles after the main outburst ended.

2. We also observed a short outburst near the end of the observing season. Thus the star appears to be a fairly normal SU UMa-type dwarf nova, except for the remarkably short $P_{\rm orb}$.

3. The superhump period exceeds the orbital period by 3.7±0.3%, which, together with the spectroscopy, indicates $q$=0.19±0.02.

4. Photometry in quiescence shows a double-humped wave with 0.022283(1) d, exactly half the orbital period. This clearly arises from the ellipsoidal distortion of the secondary. Analysis of the amplitudes constrains $q(i)$, and for the allowed $q$ we estimate $i$=50±5°. The distance is 180±40 pc.

5. Coupled with these constraints on $q$ and $i$, the radial velocities then supply a dynamical solution for the masses. We estimate $M_2$=$0.12^{+0.03}_{-0.02}\ M_\odot$, $M_1$=$0.63^{+0.12}_{-0.09}\ M_\odot$.

6. Of course, this $M_2$ makes it paradoxical that the secondary is a K star, since stars of solar composition have a spectral type M or later for masses below ~0.5 $M_\odot$. The paradox can be solved by invoking a high helium abundance in the secondary, suggesting that some considerable fraction of core H-burning occurred in the secondary before the binary reached its common-envelope stage (T02). This would occur, for example, if the original masses in the binary were nearly equal.

Since the data described in this paper were obtained with very small telescopes (averaging 35 cm), and since we caught only a portion of one long outburst, we can expect to learn a lot more about this fascinating star! We gratefully acknowledge financial support from the Research Corporation (GG-0042) and the National Science Foundation (00-98254), contributions of data by Robert Fried and James Hannon, and Patrick Schmeer for discovering this precious eruption.

TABLE 1
LOG OF OBSERVATIONS (JD 2452219–71)

| Observer | Nights/hours | Telescope |
|---|---|---|
| T. Krajci | 26/93 | CBA–New Mexico 28 cm |
| D. Skillman | 16/59 | CBA–East 66 cm |
| E. Beshore | 9/41 | CBA–Colorado 35 cm |
| D. Starkey | 8/25 | CBA–Indiana 35 cm |
| B. Martin | 6/23 | CBA–Alberta 30 cm |
| J. Kemp | 5/15 | MDM 130 cm |
| T. Vanmunster | 3/15 | CBA–Belgium 35 cm |
| R. Rea | 2/8 | CBA–Nelson 35 cm |
| A. Oksanen | 3/6 | CBA–Finland 40 cm |

TABLE 2
SIGNALS DETECTED IN POWER SPECTRUM

| Date | *V* mag | Frequencies (cycles/day) | | | |
|---|---|---|---|---|---|
| 18–21 | 13.5 | 21.57 | 43.11 | | |
| 22–24 | 15.2 | 21.64 | 43.23 | | |
| 25–27 | 15.9 | 21.78 | 43.46+44.88 | 65.22 | 86.92 |
| 27–31 | 16.1 | 21.84 | 44.88 | 65.27 | |
| 30–34 | 16.2 | 21.79 | 44.88+43.60 | | |
| 34–40 | 16.3 | 21.80 | 44.88+43.59 | | |
| 40–47 | 16.4 | (21.85) | 44.88 | | |
| 51–57 | 16.5 | | 44.88 | | |

NOTES: The error in each frequency measurement is approximately ±0.10/*N* c/d, where *N* is the duration in days. Parentheses indicate uncertain detection.

TABLE 3
TIMES OF SUPERHUMP MAXIMA

| HJD 2,452,200+ | | | | | |
|---|---|---|---|---|---|
| 19.365 | 21.590 | 23.488 | 24.508 | 25.663 | 28.510 |
| 19.410 | 21.637 | 23.534 | 24.557 | 25.710 | 29.4585 |
| 19.595 | 21.684 | 23.581 | 24.603 | 26.492 | 30.5610 |
| 19.642 | 21.730 | 23.626 | 24.648 | 26.538 | 31.3412 |
| 20.012 | 21.915 | 23.717 | 24.695 | 26.584 | 32.4865 |
| 20.058 | 21.962 | 23.766 | 24.740 | 26.631 | 33.4530 |
| 20.150 | 22.009 | 24.274 | 24.787 | 27.457 | 34.6445 |
| 20.613 | 22.473 | 24.322 | 25.479 | 27.502 | 35.5621 |
| 20.661 | 22.519 | 24.367 | 25.524 | 27.547 | 38.5520 |
| 21.498 | 22.564 | 24.414 | 25.569 | 27.593 | 40.5603 |
| 21.544 | 22.611 | 24.462 | 25.615 | 28.373 | 42.5725 |

NOTE: Times after JD 28 are consolidated to 1 per night, to improve accuracy (these runs are strongly contaminated by the orbital wave, which was subtracted to permit a timing).

## FIGURE CAPTIONS

FIGURE 1. — Eruption light curve of RX 2329+06, showing the tail end of a superoutburst, and two likely normal outbursts (with the first appearing to be an "echo" outburst). A freehand curve has been added to help trace the brightness changes. Zero magnitude on this scale corresponds to $V \cong 12.5$.

FIGURE 2. — *Upper frames*, nightly light curves on JD 21 and 25, with linear trends removed. The effective wavelength is ~6000 Å, near the *R* band. *Lowest frame*, power spectrum of JD 19–26, showing a superhump signal at 21.60 c/d and its first harmonic. Errors are ±0.02 c/d. Inset is the mean waveform, summed at this frequency (averaging over the variations in period and waveform).

FIGURE 3. — *Lower frame*, O–C diagram of superhump maxima with respect to a test ephemeris: HJD 19.361+0.046*E*. The curvature indicates a decreasing period. There is a rapid phase change of –0.38±0.07 cycles near *E*=200, possibly indicating a transition to late superhumps. *Upper frame*, the corresponding eruption light curve. Note that the superhumps outlive the main eruption by at least 400 cycles.

FIGURE 4. — Power spectrum of JD 29–68, showing the signal at 44.877±0.007 c/day, twice the orbital frequency. The signal rises off-scale to a power of 960, and the power spectrum has been "cleaned" for its aliases. Inset is the mean orbital waveform, showing the asymmetric double sinusoid arising from Roche geometry.

FIGURE 5. — The curves show $M_1$– $M_2$ solutions for a mass function of 0.194 $M_\odot$ (T02). The lines enclose the $q$=0.19±0.02 constraint obtained from superhumps and emission-line radial velocities, and the black region shows the subset of that which also satisfies the $i$=50±5° constraint (arising from the ellipsoidal variation).

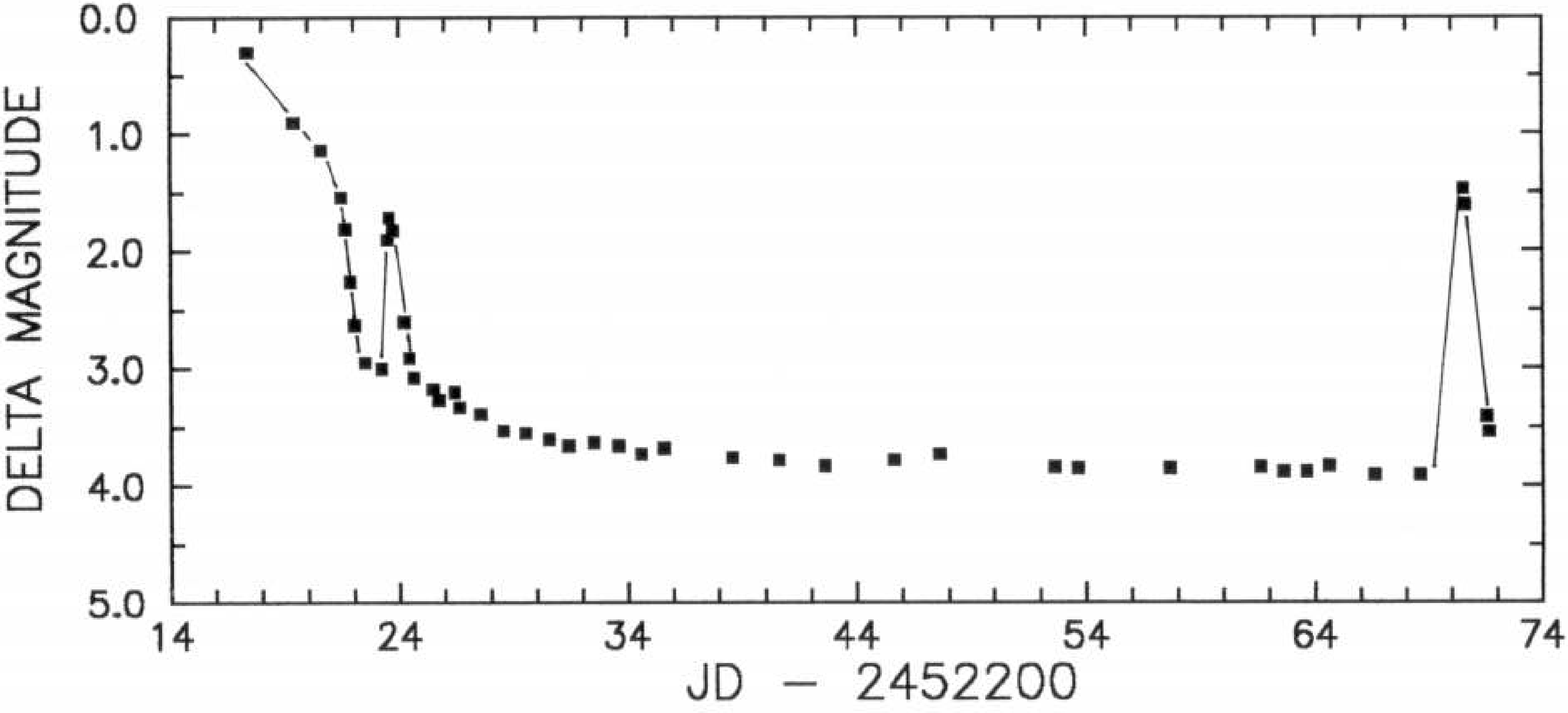

Fig 1

Fig 2

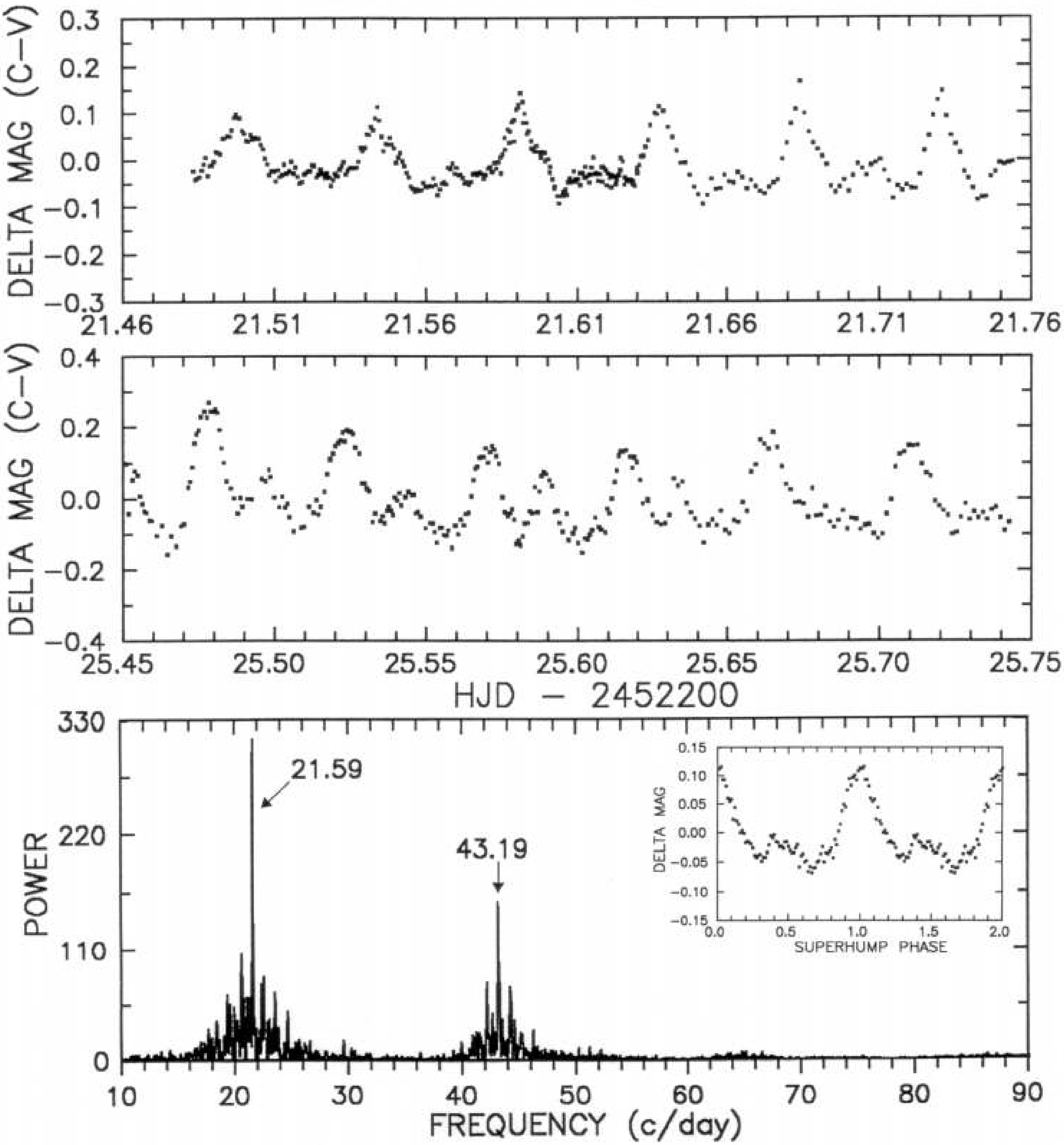

Fig 3

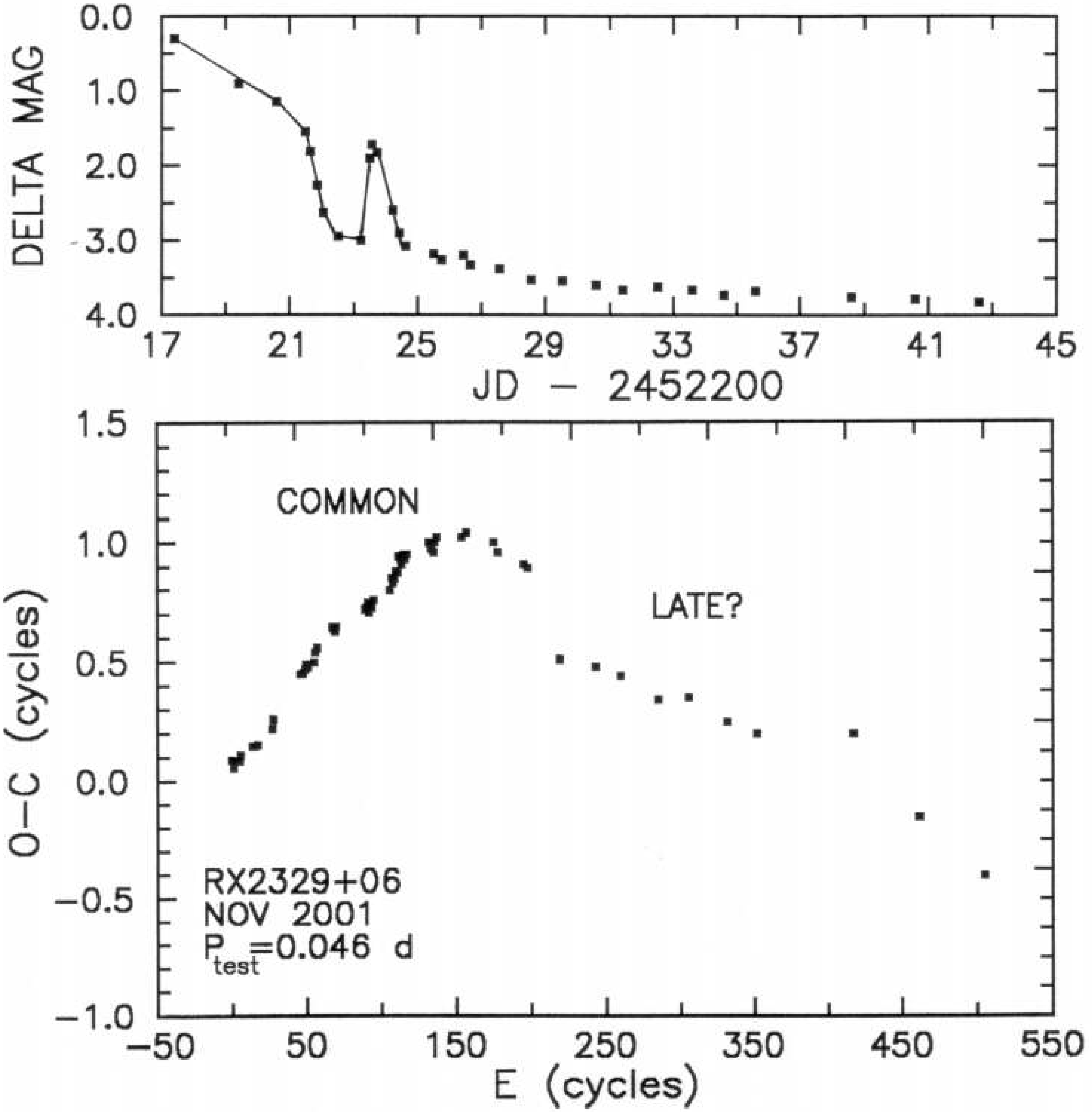

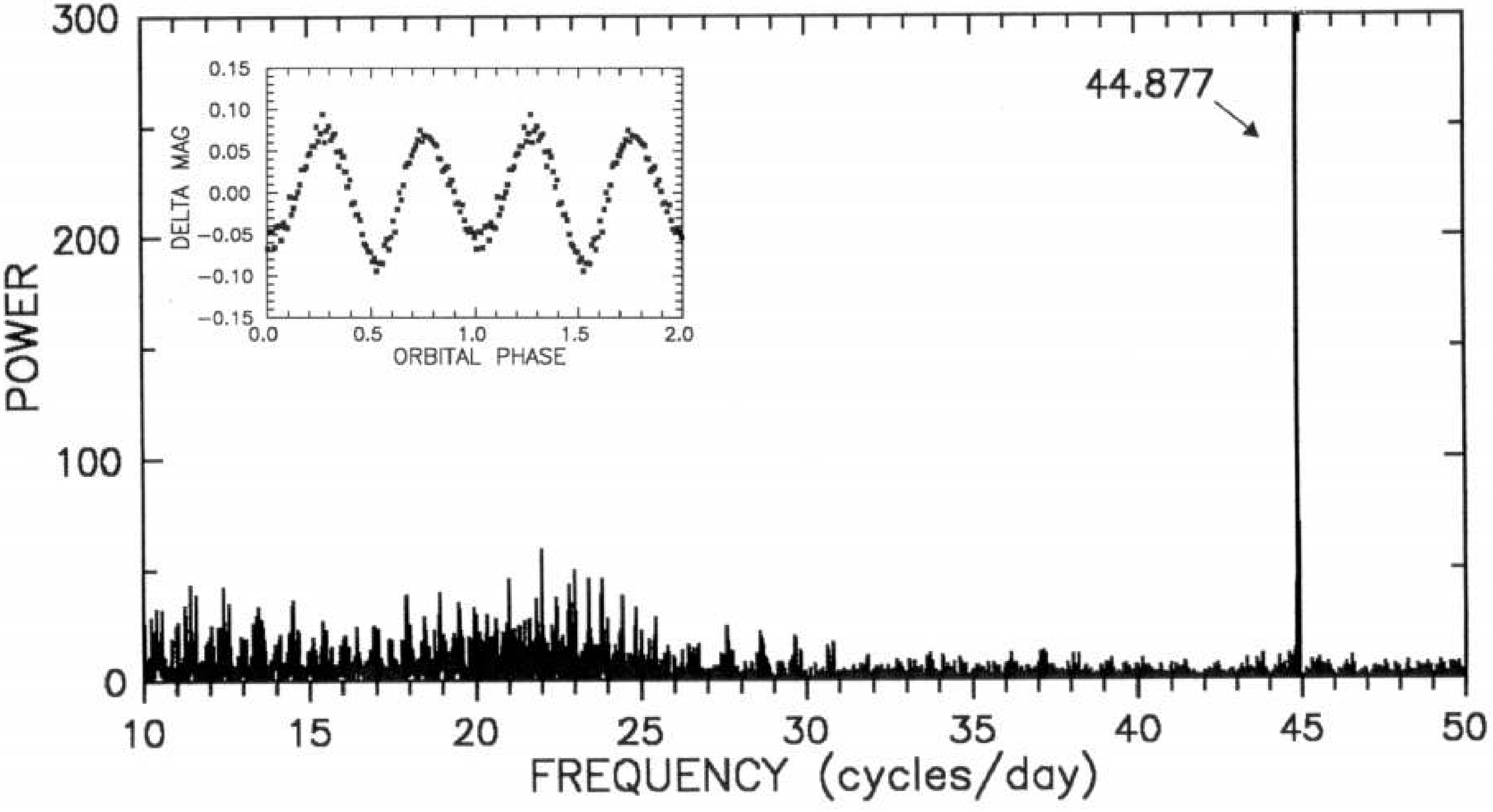

Fig 4

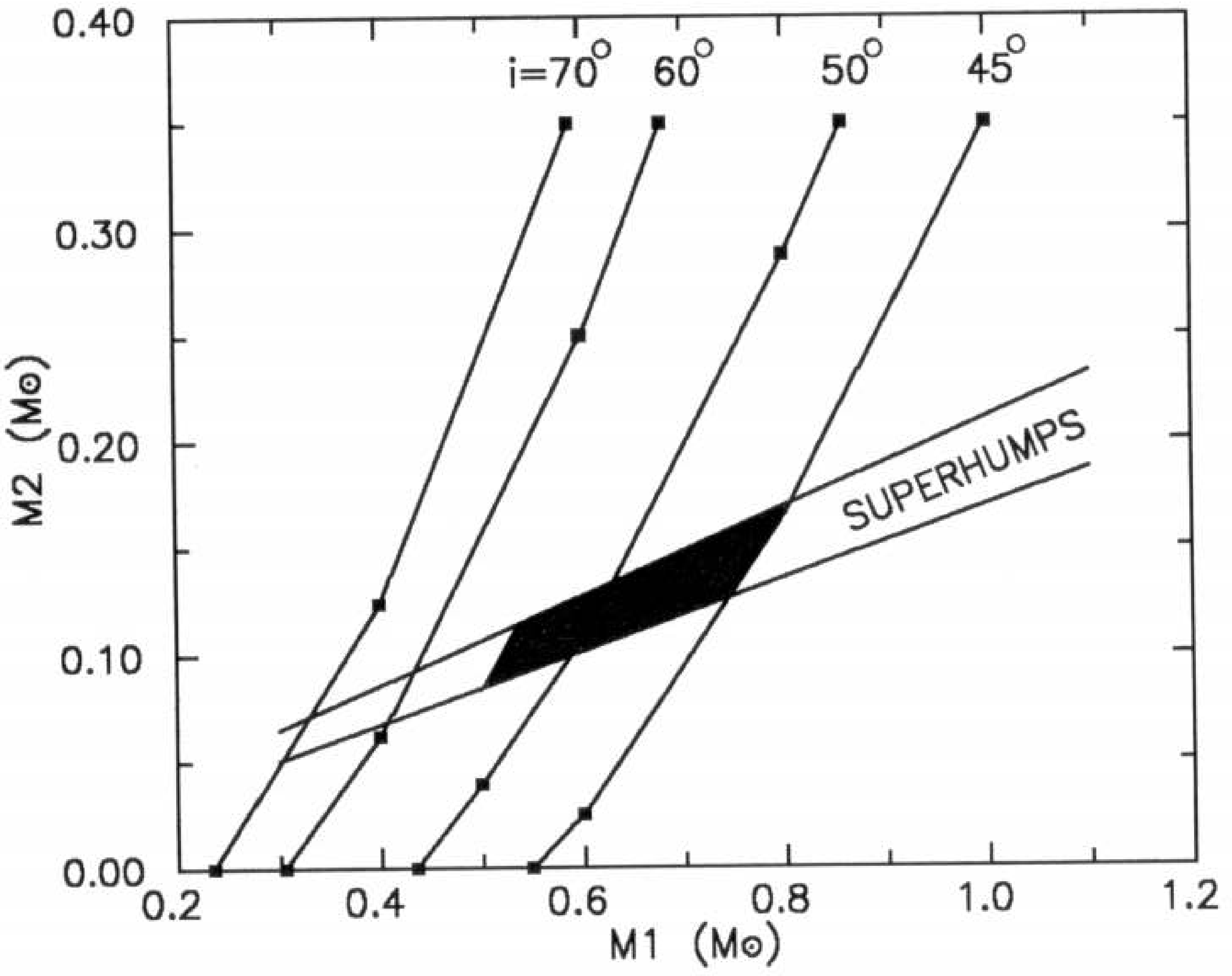

Fig 5